# Synthesis, magnetic properties and electronic structure of $S = ½$ uniform spin chain system InCuPO$_5$

B. Koteswararao,[1*] Binoy K. Hazra,[1] Dibyata Rout,[1] P. V. Srinivasarao,[2] S. Srinath,[1] S. K. Panda[3**]

[1]*School of Physics, University of Hyderabad, Central University PO, Hyderabad 500046, India.*
[2]*Department of Physics, Acharya Nagarjuna University, Nagarjuna Nagar 522 510, India.*
[3]*Centre de Physique Theorique (CPHT), Ecole Polytechnique, 91128 Palaiseau cedex, France.*

**Abstract**

We have studied the structural, magnetic properties, and electronic structure of the compound InCuPO$_5$ synthesized by solid state reaction method. The structure of InCuPO$_5$ comprises of $S = ½$ uniform spin chains formed by corner-shared CuO$_4$ units. Magnetic susceptibility ($\chi(T)$) data shows a broad maximum at about 65 K, a characteristic feature of one-dimensional (1D) magnetism. The $\chi(T)$ data is fitted to the coupled, $S = ½$ Heisenberg antiferromagnetic (HAFM) uniform chain model that gives the intra-chain coupling ($J/k_B$) between nearest neighbour Cu$^{2+}$ ions as -100 K and the ratio of inter-chain to intra-chain coupling ($J''/J$) as about 0.07. The exchange couplings estimated from the magnetic data analysis are in good agreement with the computed values from the electronic structure calculations based on density functional theory + Hubbard $U$ (DFT+U) approach. The combination of theoretical and experimental analysis confirms that InCuPO$_5$ is a candidate material for weakly coupled $S = ½$ uniform chains. A detailed theoretical analysis of the electronic structure further reveals that the system is insulating with a gap of 2.4 eV and a local moment of 0.70 $\mu_B$/Cu.

## 1. Introduction

The study of one-dimensional (1D) magnetic systems has been attracted a considerable attention in the condensed matter physics because of their exotic ground states, where the quantum fluctuations play a significant role. The interest has been initiated by the theoretical prediction that the half-integer spin chains have a gapless excitation spectrum [1], while integer-spin chains (called as Haldane chains) have a spin gap in its excitation spectrum [2]. In particular, 1D $S = ½$ Heisenberg antiferromagnetic (HAFM) systems exhibit quasi-long range ordered ground state and the elementary excitations like spinons, a domain-wall-like $S^z = ½$ excitations [3, 4]. There are two categories of $S = ½$ chains formed by direct linkage of CuO$_4$ units: by corner-shared CuO$_4$ units and by edge-shared CuO$_4$ units. A famous example for $S = ½$ spin chain with corner-shared CuO$_4$ units is Sr$_2$CuO$_3$ [5, 6]. Due to the corner-sharing picture of CuO$_4$ units, the resultant Cu-O-Cu bond angle is 180°. The super-exchange (SE) coupling ($J/k_B$) estimated from the magnetic susceptibility and heat capacity analysis is about -2200 K [6]. However, this material orders at low temperature $T_N \approx 5$ K due to the presence of inter-chain coupling [7]. Most interestingly, the separation of the spin and orbital degree of freedom (spinon and orbiton) was observed recently in Sr$_2$CuO$_3$ using resonant inelastic X-ray scattering (RIXS) experiments [8]. On the other hand, the bond-angle of Cu-O-Cu close to 90º in the spin chain due to the edge-shared picture of $CuO_4$ units favors the nearest neighbor (*nn*) interaction $J_1$ as ferromagnetic [9]. A few famous examples include $LiCuVO_4$, $Li_2CuO_2$, $Li_2CuZrO_4$, *etc.* [10-13]. In addition, some of these chains exhibit magnetic order with spontaneous electric polarization, due to the presence of frustrated next-nearest-neighbor (*nnn*) AFM coupling ($J_2$) when $|J_2/J_1|>0.25$ [14]. Recently, a few more ideal HAFM uniform chain materials, like $Sr_2Cu(PO_4)_2$ [15, 16], $Ba_2Cu(PO_4)_2$ [15, 16], $K_2CuP_2O_7$ [17], *etc.*, were reported. The chains in these categories of materials are built neither through direct edge-shared nor corner-shared $CuO_4$ units, but the coupling is rather mediated through $PO_4$ units *via* super-super-exchange (SSE) path. However, the doping ($S = 0$ or hole) effects on these materials were not studied, probably due to the reason that the SSE path might not be favorable to study the impurity-induced magnetism. A fewer materials only exist with $S = ½$ 1D HAFM uniform chains with corner-shared $CuO_4$ units. Herein, we present a quantum spin chain system $InCuPO_5$ belongs to this rare category.

In this paper, we report the details of synthesis, structural, magnetic properties and electronic structure of a new $S = ½$ uniform chain material $InCuPO_5$ [18]. The chains are formed *via* the direct corner-shared $CuO_4$ units with a bond angle of Cu-O-Cu of about 113.7º. The magnetic susceptibility shows a broad maximum at around 65 K, indicating the low-dimensional nature of this system. The antiferromagnetic SE coupling ($J/k_B$) between Cu atoms in the uniform spin chain is estimated to be about -100 K, which is in good agreement with the value extracted from the theoretical calculations, based on local spin density approximation + Hubbard $U$ (LSDA+U) method. The appearance of inter-chain coupling ($J'/J = 0.07$) suggests that $InCuPO5$ is a weakly coupled spin chain. We also studied the electronic structure of its lowest energy state and found that the system is insulating in nature.

## 2. Experimental section
### 2.1 Experimental details

The polycrystalline samples of $InCuPO_5$ were prepared using solid-state reaction method using the high purity chemicals of $In_2O_3$, $CuO$, and $(NH_4)_2HPO_4$. X-ray diffraction (XRD) data were collected on a Bruker *D*8 Focus Cu-$K_\alpha$ radiation at room temperature in the $2\theta$ range of 8–90°. The magnetic moment measurements were performed using a vibrating sample magnetometer (VSM) option attached to Physical Property Measurement System (PPMS), Quantum Design, *Inc.*, in an applied magnetic field ($\mu_0 H$) of 1 T and in the temperature ($T$) range 10 to 300 K.

### 2.2 Synthesis

In the first attempt, we have tried to prepare the title compound using the chemicals $In_2O_3$, $CuO$, and $(NH_4)_2HPO_4$ with the exact stoichiometry of the elemental molar ratio 0.5:1:1. The mixture was grinded and fired at different steps 250°C, 350°C, 500°C for 6 hours each with intermediate grindings in order to decompose $NH_3$ and $H_2O$ from $(NH_4)_2HPO_4$ chemical. We have fired at 850°C for 12 hrs and finally at 950°C for 12 hours to get InCuPO$_5$ compound phase. In this attempt, we have found an inevitable, large impurity phase of In$_2$Cu$_3$(PO$_4$)$_4$ [19], which we labeled as sample-1. The relative height of the impurity peak is about 20 % when compared to that of the maximum XRD peak (0 1 4) of original phase InCuPO$_5$ (see inset of Fig. 1). The formation of impurity phase might be due to the volatizing nature of InPO$_4$ during the firing. To reduce the impurity phase in the XRD, we have added 10 % of InPO$_4$ to the above chemicals. Finally we reduced the height of the impurity phase to 7 %, as shown in the inset of Fig. 1. We have also attempted the optimization of this sample preparation by adding the different concentrations of InPO$_4$, but the sample batch with an additional concentration of 10% InPO$_4$ was found to be a better among all our attempts. Therefore, we used the latter batch (sample-2) for the magnetic analysis in this paper.

### 2.3 Computational details

In order to understand the magnetic behavior and the electronic structure of InCuPO$_5$, we have carried out first-principles density functional theory calculations using the Vienna *ab*-initio simulation package (VASP) code [20, 21] within the projector augmented wave (PAW) method [22]. The exchange and correlation effects are treated using the local spin density approximation (LSDA) as well as including Hubbard *U* within the LSDA+U framework [23]. The LSDA+U calculations are carried out for two possible values of on-site Coulomb interaction of Cu-d states namely $U = 6$ and 8 eV, while the on-site exchange parameter was fixed at 0.8 eV. The kinetic energy cut off of the plane wave basis was chosen to be 650 eV. Brillouin-zone integration was performed using a 15x17x15 k-mesh. All the calculations are carried out with the experimental structure, reported in this study.

### 3. Results and discussion
### 3.1 X-ray diffraction and structural details

Fig. 1 shows the X-Ray diffraction (XRD) plot of InCuPO$_5$. We have employed the Rietveld refinement using Fullprof software [24]. The lattice constants are obtained to be $a = 7.750$ (5) Å, $b = 6.615$ (5) Å, and $c = 7.314$ (5) Å, which are comparable to the reported values of InCuPO$_5$ unit cell [18]. The refined lattice parameters are mentioned in Table I.

The compound InCuPO$_5$ crystallizes in an orthorhombic unit cell with a space group *Pnma* (Space group number 62) [18]. As shown in Fig. 2(a), the crystal structure of InCuPO$_5$ is formed by the building blocks of CuO$_4$ square units, PO$_4$ tetrahedral units, and In atoms. The basic units of CuO$_4$ and PO$_4$ are shown in Fig. 2(b). The details of bond lengths and bond angles are mentioned in the Table II and III. In this structure, the CuO$_4$ square planar units are connected each other through their corners, forming Cu-O-Cu chains propagating along the crystallographic *b*-direction (see Fig. 2(c)). Similar corner-shared CuO$_4$ units forms a famous $S = ½$ 1D HAFM uniform chain material Sr$_2$CuO$_3$. The bond angle of Cu-O4-Cu in the InCuPO$_5$ is about 113.7°. In such a crystal geometry, the SE coupling between Cu$^{2+}$ ions in the spin chain is expected to be antiferromagnetic according to Goodenough predictions [9]. These spin chains are well separated from each other by the presence of $In^{3+}$ ions between the chains. A weak inter-chain coupling is expected through $PO_4$ units.

### 3.2 Magnetic susceptibility analysis

The magnetic susceptibility $\chi(T)$ data show a Curie-Weiss behavior at high-temperature. A broad maximum-like feature is seen at 65 K in the $\chi(T)$ data, a characteristic feature of low-dimensional magnetism attributed to short range correlations. At low-$T$, the data increases again like Curie-Weiss behavior. As the structure suggests that it has uniform chains formed due to corner-shared $CuO_4$ units (see Fig. 2 (c)) and also as the XRD confirms the presence of a magnetic impurity phase, we have fitted the measured $\chi(T)$ data with the following expression for a coupled, $S = ½$ HAFM uniform spin chain model [6] in the temperature range from 10 to 300 K.

$$\chi(T) = \chi_0 + \frac{C_{imp}}{(T - \theta_{imp})} + \chi(J, J', T) \quad, \text{where } \chi(J, J', T) = \frac{\chi_{chain}(J, T)}{1 + \frac{zJ'}{Ng^2\mu_B^2}\chi_{chain}(J, T)}$$

From the above fitting, we obtained the value of temperature independent susceptibility $\chi_o$ as -4.18×10$^{-4}$ cm$^3$/mol. The parameters of Curie constant ($C_{imp}$) and Curie-Weiss temperature ($\theta_{CW}$) account the intrinsic and extrinsic impurity contributions. These are found to be $C_{imp}$ = 0.1 cm$^3$ K/mol and $\theta_{CW}$ = -15 K, which are somewhat large owing to the presence of magnetic impurity phase of $In_2Cu_3(PO_4)_4$ in our material. The above fitting provides a value of intra-chain coupling $(J/k_B) = - 100 \pm 3$ K, which is in good agreement with the expected value of broad maximum for a 1D spin-chain magnet *i.e.*, 0.640851 $J/k_B$ [6]. The total strength of inter-chain coupling z$J'$ (z being the total number of nearest neighbor Cu chains) are found to be (27±1) K. Since each chain in InCuPO$_5$ has four nearest-neighbor chains *i.e.*, z = 4, the ratio of inter-chain to intra-chain coupling ($J'/J$) is estimated to be about 0.07.

To further understand uniform spin chain behavior, we have subtracted $\chi_o$ and impurity contribution $C_{imp}/(T-\theta_{CW})$ from the measured magnetic susceptibility $\chi(T)$ to get the intrinsic spin chain susceptibility $\chi_{chain}$. The normalized spin chain susceptibility *i.e.*, $\chi_{chain}$ divided by $Ng^2\mu_B^2/J$ is plotted as a function of normalized temperature (k$_B T/J$), as shown in Fig. 4. Here $N$, $g$, $\mu_B$, and k$_B$ are the Avogadro number, Lande-g factor, Bohr magneton, and Boltzmann constant, respectively. The normalized spin chain susceptibility shows the maximum value of 0.14 at k$_B T^{max}/J = 0.65$, which is in good agreement with the theoretically expected value 0.146 for $S = ½$ 1D HAFM uniform chain model [6]. We note that the magnetic long-range order is expected at finite-*T* based on our *J* and z*J*´values. The relation between $T_N$, z*J*´ and *J* is as follows [6].

$$zJ' = \frac{T_N}{0.2333\sqrt{\ln\left(\frac{2.6J}{T_N}\right)+\frac{1}{2}\ln\left(\ln\left(\frac{2.6J}{T_N}\right)\right)}}$$

According the above formula, the $T_N$ is expected to be appeared at 11 K. However, we did not observe any anomaly in our magnetic data. This is probably due to the large Curie-Weiss upturn in χ(*T*) at low temperatures due to the presence of magnetic impurity phase $In_2Cu_3(PO_4)_4$ in this material, which hides the intrinsic behavior at low-*T*. Local probe measurements such as NMR and μSR and single crystalline samples are warranted to probe the further physics of this material. In the following section, we will corroborate our experimental results with the first principles electronic structure calculations.

### 3.3 Electronic structure

In order to analyze the electronic structure of $InCuPO_5$ and to provide a detailed microscopic understanding of our experimental results, we have carried out first principles calculations within LSDA and LSDA+U approaches. At first to investigate the lowest energy magnetic state, we have simulated four different magnetic configurations, namely FM (both intra-chain and inter-chain couplings are ferromagnetic), AFM1 (intra-chain couplings are antiferromagnetic and inter-chain couplings are ferromagnetic), AFM2 (intra-chain couplings are ferromagnetic and inter-chain couplings are antiferromagnetic), and AFM3 (both intra-chain and inter-chain couplings are antiferromagnetic) as shown in Fig. 5. Our results of LSDA and LSDA+U calculations as summarized in Table IV reveal that the AFM3 configuration is the lowest energy state within both the approaches. As expected the spin moment on the Cu site is smaller in LSDA and it increases with increasing the value of *U*. Table IV also reveals that the Cu magnetic moments remain almost unaltered for different magnetic configurations, implying the localized nature of the Cu ions. In view of that, a localized picture based on Heisenberg model would be a reliable option for the description of the magnetic exchange couplings in $InCuPO_5$. To quantify the exchange interactions, we assume that the energy differences between the different magnetic states are predominantly contributed from the interaction between the Cu moments with spin **S**, which can be effectively modeled by the following Heisenberg model.

$$H = J\sum S_i \cdot S_j + J'\sum S_i \cdot S_j$$

Where, *J* and *J*´ are the intra-chain and inter-chain coupling between the Cu ions as marked in Fig. [2]. The exchange interactions can be obtained by mapping the computed total energies corresponding to various magnetic states (Table IV) onto the corresponding total spin exchange energies of the above

Heisenberg Hamiltonian. The computed exchange parameters following such procedure are displayed in Table [V]. We find that the magnitudes of the exchanges are highest in LSDA and decrease with increasing the value of *U*. This is in accordance with the Kugel-Khomskii model [25, 26] where magnetic exchanges *J* are inversely proportional to the magnitude of *U*. Our estimated *J*'s (Table V) clearly reveal that the only significant exchange is the *nn* intra-chain interaction and that is AFM in nature. The *nnn* inter-chain interaction (*J'*) is considerably small for all the three cases and it is about 6.5% of *J* for both *U* = 6 and 8 eV. This is an excellent agreement with the experimental data analysis. According to the structure, the inter-chain coupling *J'* is attributed to the exchange path between $CuO_4$ via $PO_4$ units. In the $PO_4$ tetrahedral unit, the bond length of P-O1 is 1.652 Å which is larger compared to the other bond-lengths P-O2=1.528 Å and P-O3=1.510 (see Table III). Since the *J'* is mediated *via* the longer bond length of P-O3 in the exchange path Cu-O1-P-O3-Cu, the inter-chain coupling is relatively weaker compared to that of *J* which is mediated through direct Cu-O-Cu path. However, *J'* would play a significant role in establishing the long range order at finite temperature, which should be further confirmed by other local probe measurements. Since the correlation strength is usually very high for Cu-based oxide materials [27, 28], the computed exchanges for *U* = 8 eV can be considered as the values of exchanges in this material. Thus our computational exchange interactions further provide credence to the measured magnetic susceptibility data analysis, concluding that $InCuPO_5$ is a member of weakly coupled *S* = ½ HAFM uniform chain system. The spin moment on the Cu ion at *U* = 8 eV is 0.70 $\mu_B$. This is also in good agreement with the reported Cu moments on other Cu-based materials [29, 30].

Next we discuss the electronic structure of $InCuPO_5$ in its lowest energy state (AMF3). The partial density of states of the Cu-*d* states for both the spin channels are displayed in Fig. 6. Our result shows that the majority states are completely occupied and the gap opens up in the minority channel. This is expected since Cu is in nominal 2+ charge state ($d^9$), the first 5 electrons occupy the majority channel, while the rest 4 electrons fill up the minority states, keeping only one level empty. Since AFM3 is symmetry broken phase, a gap open up between the two spin-states of the top most half-filled *d* level. The value of the gap is smallest in LSDA and on-site Coulomb correlation effect in form of *U* within LSDA+U approach further localize the occupied states, increasing the distance between the occupied and the un-occupied levels (see Fig. 6). Thus insulating gap increases with increasing the value of *U* and it is 2.4 eV at U=8 eV.

### 4. Conclusion

We have prepared the polycrystalline samples of a new *S* = ½ uniform chain material $InCuPO_5$ using solid state reaction method by adding extra 10% concentration of $InPO_4$ to reduce the formation of impurity phases. Magnetic susceptibility data show a broad maximum at about 65 K due to the short range correlations, indicating the low-dimensional nature of the magnetic material. Magnetic data analysis give the coupling between *nn* spins in the uniform chain is antiferromagnetic as expected from bond-angle of Cu-O-Cu about 113.7°. The values of the intra- and inter-chain exchange coupling obtained from $\chi(T)$ are about $J/k_B$ = -100 K and $J' = 0.07 J$, which are in good agreement with the values estimated from DFT+U based calculations. In summary, both our experimental and theoretical analyses establish that InCuPO$_5$ belongs to the class of coupled, 1D $S = ½$ Heisenberg antiferromagnetic spin-chain insulating system, formed by corner-shared CuO$_4$ units. Further, local probe techniques $^{31}$P NMR, μSR, *etc.,* would be helpful to have more detailed understanding of the magnetic ground state properties and spin dynamics of this new uniform chain material.

**Acknowledgements:** B.K. thanks the funding from DST INSPIRE faculty award-2014 scheme.
* koti.iitb@gmail.com
** swarup.panda@polytechnique.edu

**Figures and Captions:**

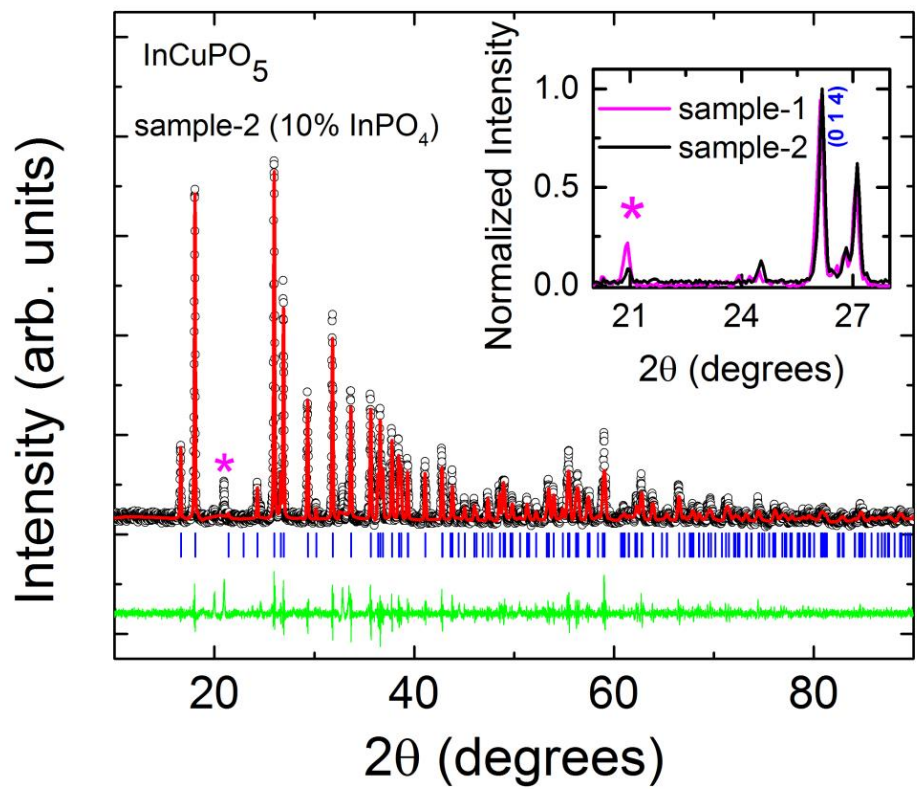

Fig. 1. X-ray diffraction of InCuPO$_5$ polycrystalline samples with Rietveld refinement analysis for the sample-2. The residual parameters of Rietveld refinement are R$_p$ ≈ 36.5 %, R$_{wp}$ ≈ 32.5 %, R$_{exp}$ ≈ 10.5 %, and $\chi^2$ ≈ 6.83, respectively. The * mark indicates the impurity contribution from In$_2$Cu$_3$(PO$_4$)$_4$ phase. Inset shows the comparison between the XRD of two batches of sample-1 (prepared with stoichiometric chemicals) and sample-2 (prepared with addition of 10% InPO$_4$).

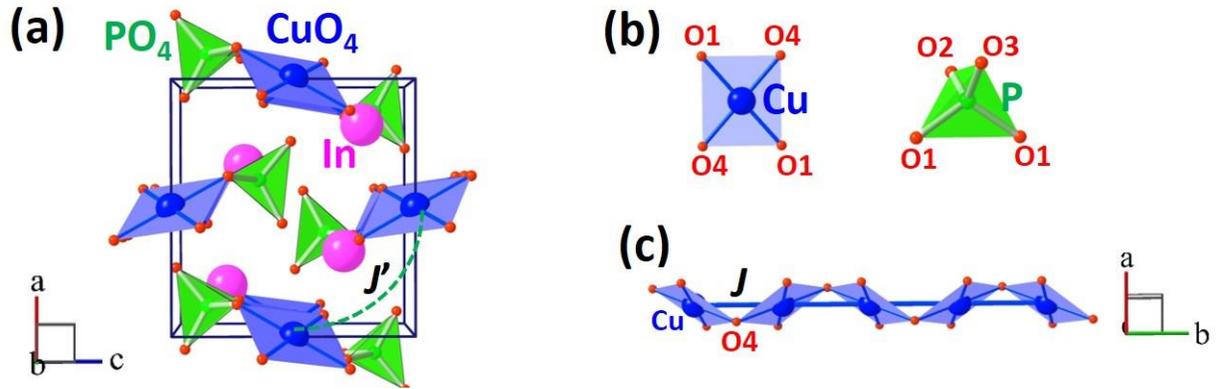

Fig. 2. (Color online). (a) Crystal structure of InCuPO$_5$. (b) The environments of CuO$_4$ square plane and PO$_4$ tetrahedral units. (c) The uniform chain passing along the crystallographic $b$-direction.

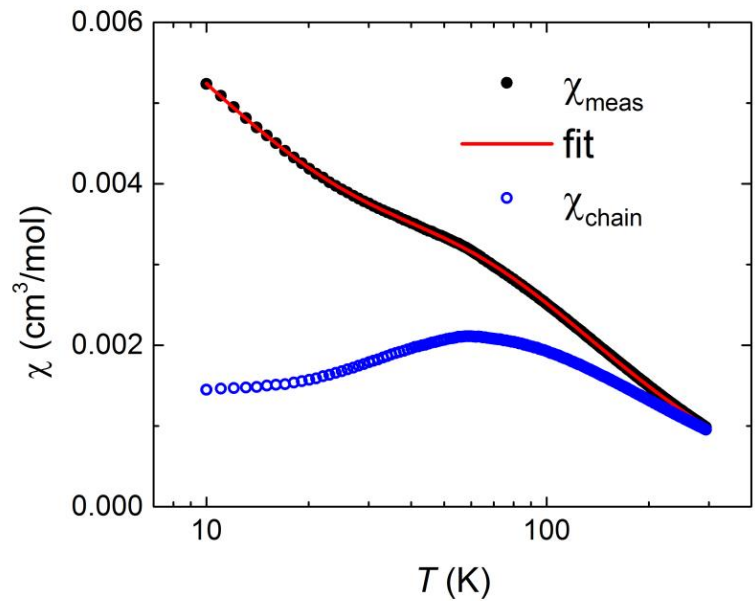

Fig. 3 (color online). (a) Magnetic susceptibility $\chi$ is plotted in the temperature range 10 – 300 K. The red line is the fit to $S = \frac{1}{2}$ HAFM uniform chain model (red solid line). Open circles (blue) indicate the intrinsic uniform chain susceptibility ($\chi_{chain}$) data after the subtraction of impurity contribution.

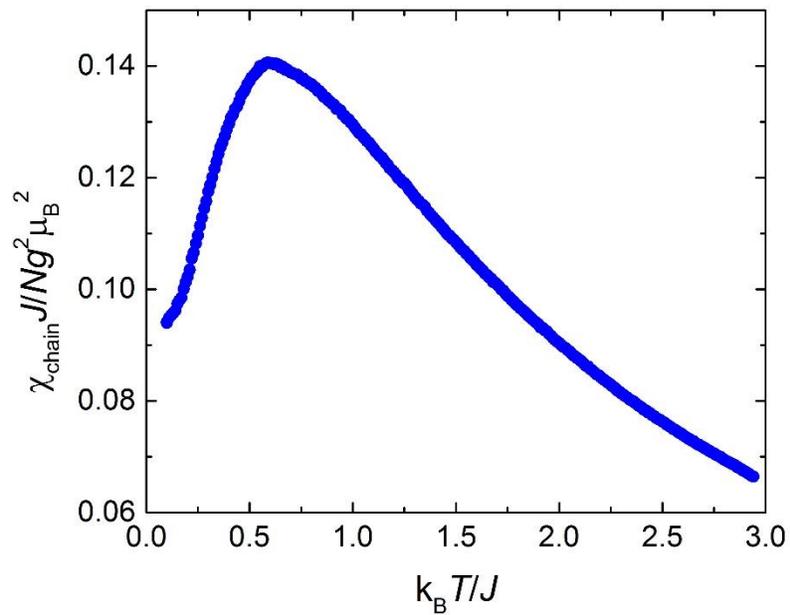

Fig. 4 (color online) Normalized magnetic susceptibility versus normalized temperature.

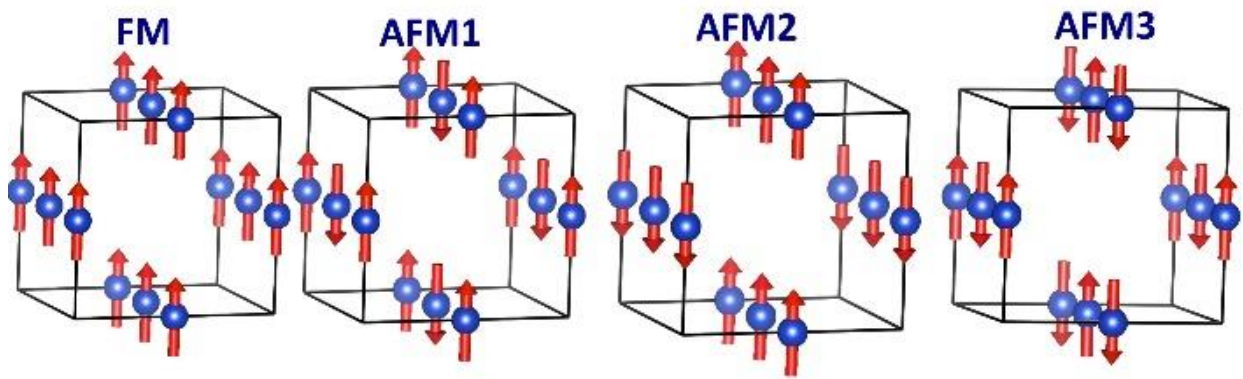

Fig.5. Schematic picture of a few possible magnetic configurations in $InCuPO_5$ where only the Cu spins are displayed.

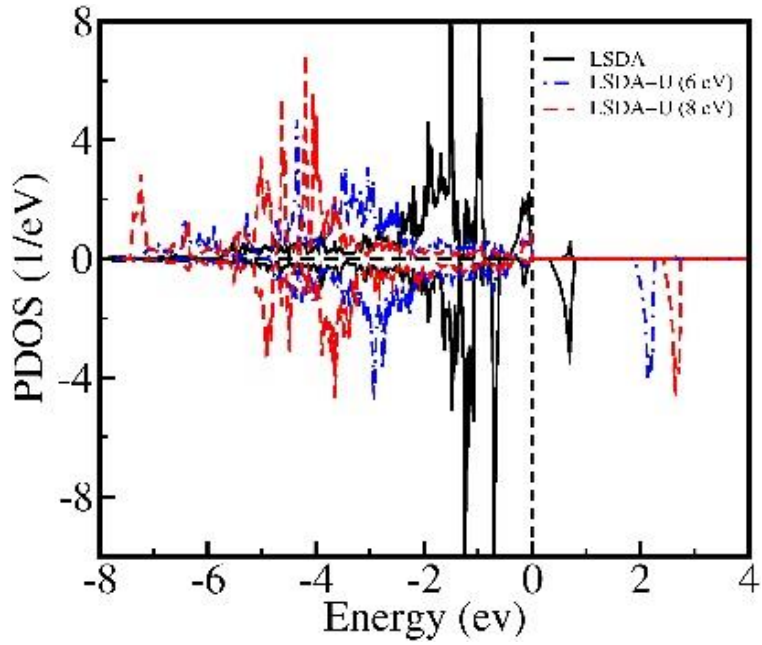

Fig. 6 (color online) Partial density of states (PDOS) of the Cu-d states in InCuPO$_5$ within LSDA and LSDA+U approaches.

## Tables and Captions

Table 1. The details of atomic coordinates and occupancies in the spin chain InCuPO$_5$.

| Atoms | Wyckoff | x | Y | Z | occupancy |
|---|---|---|---|---|---|
| Cu | *4b* | 0 | 0 | 0.5 | 1 |
| In | *4c* | 0.174 | 0.25 | 0.212 | 1 |
| P | *4c* | 0.115 | 0.75 | 0.150 | 1 |
| O1 | *8d* | 0.124 | 0.957 | 0.269 | 1 |
| O2 | *4c* | 0.064 | 0.25 | 0.954 | 1 |

| O3 | *4c* | 0.239 | 0.25 | 0.510 | 1 |
| O4 | *4c* | 0.430 | 0.25 | 0.127 | 1 |

Table II. The details of bond lengths in the crystal structure of InCuPO$_5$

|  | Bonds | Bond length (Å) |
|---|---|---|
| CuO$_4$ | Cu-O1 | 1.964 |
|  | Cu-O4 | 1.973 |
| PO$_4$ | P-O1 | 1.652 |
|  | P-O2 | 1.528 |
|  | P-O3 | 1.510 |
| InO$_8$ | In-O1 | 2.022 |
|  | In-O2 | 2.065 |
|  | In-O3 | 2.234 |
|  | In-O4 | 2.073 |

Table III. The details of bond lengths and bond angles related to the magnetic couplings in InCuPO$_5$

| Cu-Cu bond length (Å) | Bond angle (º) |
|---|---|
| 3.30 | Cu-O4-Cu = 113.7 |
| 5.32 | Cu-O1-P=124.7 |
|  | P-O3-Cu=126.4 |
| 6.26 | Cu-O1-P=124.7 |
|  | P-O3-Cu=126.4 |

Table IV: Energies ΔE (meV/Cu) for different magnetic ordering with respect to ferromagnetic alignment between Cu spins, as well as spin moment $\mu_s$ ($\mu_B$/Cu) of Cu$^{2+}$ ions.

|  | LSDA |  | LSDA+U (6 eV) |  | LSDA+U (8 eV) |  |
|---|---|---|---|---|---|---|
|  | ΔE | $\mu_s$ | ΔE | $\mu_s$ | ΔE | $\mu_s$ |
| FM | 0.00 | 0.59 | 0.00 | 0.67 | 0.00 | 0.71 |
| AFM1 | -24.51 | 0.50 | -6.36 | 0.65 | -4.31 | 0.70 |
| AFM2 | -2.63 | 0.57 | -0.83 | 0.66 | -0.56 | 0.71 |
| AFM3 | -27.28 | 0.51 | -7.43 | 0.65 | -5.06 | 0.70 |

Table V: Intra-chain ($J$) and inter-chain ($J'$) magnetic couplings between the $Cu^{2+}$ ions in $InCuPO_5$. The corresponding distances are also given. Negative sign indicate the antiferromagnetic nature of the exchange.

|  | $d_{Cu-Cu}$ (Å) | LSDA | LSDA+U (6 eV) | LSDA+U (8 eV) |
|---|---|---|---|---|
| $J$ (meV) | 3.30 | -49.02 | -12.72 | -8.62 |
| $J'$ (meV) | 5.32 | -2.63 | -0.83 | -0.56 |